\documentclass[printer]{aa}
\usepackage{psfig}

\def\s5{S5 2007+777}
\def\3c{3C371}
\def\etal{et al.}

\begin{document}
\title{B and I-band optical micro-variability observations of the BL Lac
objects S5 2007+777 and 3C371}
\author{E.M. Xilouris\inst{1} \and I.E. Papadakis\inst{2} \and P. Boumis 
\inst{1} \and A. Dapergolas\inst{1} \and J. Alikakos\inst{1}
\and J. Papamastorakis\inst{2,3} \and N. Smith\inst{4} \and 
C.D. Goudis\inst{1,5} }
\offprints{E.M. Xilouris;  e-mail: xilouris@astro.noa.gr}
\institute{Institute of Astronomy \&
Astrophysics, National Observatory of Athens, I. Metaxa \& V. Pavlou, P.
Penteli, GR-15236, Athens, Greece 
\and Physics Department, University of Crete, P.O. Box 2208,
   GR-71003, Heraklion, Crete, Greece
\and IESL, Foundation for Research and Technology-Hellas, P.O.Box
1527, GR-71110, Heraklion, Crete, Greece
\and Department of Applied Physics \& Instrumentation, Cork Institute of Technology,
Cork, Ireland
\and Department of Physics, University of Patras, GR-26500, Rio-Patras, Greece}

\date{Received... ; accepted... }
\abstract{We have observed  \s5\ and \3c\ in the $B$ and $I$ bands for 13 and 8
nights, respectively, during various observing runs in 2001, 2002 and 2004. 
The observations resulted in almost evenly sampled light curves, $6-9$ hours
long. We do not detect any flares within the observed light curves, but we do
observe small amplitude, significant variations, in both bands, on time scales
of hours and days. The average variability amplitude on time scales of
minutes/hours is $\sim 2.5$\% and $\sim 1-1.5$\% in the case of \s5\ and \3c,
respectively. The average amplitudes increase to $\sim 5-12$\% and $\sim
4-6$\%, respectively, on time scales of days.  We find that the $B$ and $I$
band  variations are highly correlated, on both short and long time scales. 
During the 2004 observations, which resulted in the longest light curves, we
observe two well defined flux-decay and rising trends in the light curves of
both objects. When the flux decays, we observe significant delays, with the $B$
band flux decaying faster than the flux in the $I$ band. As a result, we also
observe significant, flux related spectral variations as well. The
flux-spectral relation is rather complicated, with loop-like structures forming
during the flux evolution. The presence of spectral variations imply that the
observed variability is not caused by geometric effects. On the other hand, 
our results are fully consistent with the hypothesis that the observed
variations are caused by perturbations which affect different regions in the
jet of the sources.
}

\titlerunning{Optical micro-variability of S5 2007+777 and 3C371}
\authorrunning{Xilouris \etal}
\maketitle
   
%%%%%%%%%%%%%%%%%%%%%%%%%%%%%%%%%%%%%%%%%%%%%%%%%%%%%%%%%%%%%%
\section{Introduction}
%%%%%%%%%%%%%%%%%%%%%%%%%%%%%%%%%%%%%%%%%%%%%%%%%%%%%%%%%%%%%%
\smallskip

BL Lac objects constitute a class of active galactic nuclei whose optical
spectra usually do not exhibit strong emission or absorption lines. They show
high polarization and continuum variability at all wavelengths at which they
have been observed, from X-rays to radio.  Their overall spectral energy
distribution  shows two distinct components in the $\nu-\nu F_{\nu}$
representation, peaking from mm to the X--rays and at GeV--TeV energies,
respectively (e.g. Fossati \etal. 1998). The commonly accepted scenario assumes
that the non-thermal emission from BL Lacs is synchrotron and inverse-Compton
radiation produced by relativistic electrons in a jet oriented close to the
line of sight.

The galaxy 3C371 is classified as an intermediate source between BL Lacertae
objects and radio galaxies (Miller 1975). Nilsson \etal\ (1997) and Pesce
\etal\ (2001) have reported the detection of an optical and X-ray jet,
respectively, associated with the source. \3c\ was one of the first quasars
found to be variable on time scales of years/months/days. Optical variability
studies have been conducted by Oke (1967), Sandage (1967), Usher \& Manley
(1968), Miller \& McGimsey (1978), and Webb \etal\ (1998). Carini et al. (1998)
presented dense optical monitoring observations of the source in the $V$ band,
over a period of five nights. They observed intra-night variations with of the
order of $\sim 0.03$ mag hr$^{-1}$. 

\s5\ was classified as a BL Lac object by Biermann \etal\ (1981) because of its
featureless optical spectrum at the time of the observation. Wagner \etal\
(1990) did not detect any significant variations in the optical ($R$ band) and
radio bands, during observations over a period of a few nights. Recently, Peng
\etal\ (2000) presented contemporaneous radio, mm, infra-red and optical ($R$
band) observations of the source over a period of three weeks. The source was
variable in all bands on time scales of the order of a few days. 

In this work, we present simultaneous $B$ and $I$ band monitoring observations
of both \3c\ and \s5, obtained from Skinakas Observatory, Crete, Greece over a
period of 8 and 13 nights, respectively, in 2001, 2002 and 2004. The quality of
the light curves is similar to those presented by Papadakis \etal\ (2003, 2004)
in the case of BL Lac and  S4 0954+658. The present observations offer detailed
light curves, in two different bands,  to study the intra-night flux and energy
spectral variations of the two sources.

Our main aim is to use them in order to investigate the flux and spectral
variations of the source on time scales as short as a few minutes/hours. We
detect small amplitude variations that are associated with energy spectral
variations in a a rather complex way. Our results support the idea that the
fast, optical intra-night variations in BL Lac objects are mainly caused by
perturbations which activate different emitting regions in the jet. 

%%%%%%%%%%%%%%%%%%%%%%%%%%%%%%%%%%%%%%%%%%%%%%%%%%%% 
\section{Observations and data reduction} 
%%%%%%%%%%%%%%%%%%%%%%%%%%%%%%%%%%%%%%%%%%%%%%%%%%%%%

\3c\ was observed for 3 nights in 2001 and 5 nights in 2004 with the 1.3 m,
f/7.7 Ritchey-Cretien telescope at Skinakas Observatory in Crete, Greece. The
observations were carried out through the standard Johnson $B$ and Cousins $I$
filters. The CCD used was a $1024 \times 1024$ SITe chip with a 24 $\mu$m$^{2}$
pixel size (corresponding to $0^{\prime\prime}.5$ on the sky). The exposure
time per frame was 300 and 120 s for the $B$ and $I$ filters, respectively. 
\s5\ was observed for 3 nights in 2001, 4 nights in 2002 and 6 nights in 2004
with the  same telescope and camera configuration as in the case of \3c. The
exposure time per frame was 540 and 240 s for the $B$ and $I$ filters,
respectively. 

In Table 1 we list the observation dates and the number of frames that we
obtained each night for both \3c and \s5. We observed \s5\ 325 and 316 times in
the {\it B} and {\it I} bands, respectively, during the 13 nights of
observations, and \3c\ a total of 263 and 266 times, over a period of 8 nights,
in the same bands.

During the observations, the seeing varied between $\sim 1^{\prime\prime} -
2^{\prime\prime}$. Standard image processing (bias subtraction and flat
fielding using twilight-sky exposures) was applied to all frames
using standard IRAF\footnote{
IRAF is distributed by the National Optical Astronomy Observatories,
which are operated by the Association of Universities for Research
in Astronomy, Inc., under cooperative agreement with the National
Science Foundation.} routines.
For both \3c and \s5 we performed aperture photometry by integrating counts within a
circular aperture centered on the objects. For \3c we used an aperture of
$7^{\prime\prime}.5$ and for \s5 an aperture of $5^{\prime\prime}$. 

%%%%%%%%%%%%%%%%%%%%%%%%%%%%%%% TABLE 1 %%%%%%%%%%%%%%%%%%%%%%%%%%%%%%
\begin{table}
\begin{center}
\caption{Date of observations, and number of frames (nof) obtained at
each optical band.}
\begin{tabular}{lcccc} \hline
Date & Galaxy & $B$    &  $I$ \\
     &        & (nof)  & (nof) \\
\hline
14/05/01 & \s5\ & 20 & 19 \\
15/05/01 & \s5\ & 28 & 16 \\
16/05/01 & \s5\ & 29 & 27 \\
31/08/01 & \3c\ & 30 & 31 \\
01/09/01 & \3c\ & 29 & 33 \\
05/09/01 & \3c\ & 17 & 16 \\
18/09/02 & \s5\ & 20 & 21 \\
19/09/02 & \s5\ & 16 & 18 \\
20/09/02 & \s5\ & 13 & 12 \\
21/09/02 & \s5\ & 30 & 30 \\
28/07/04 & \3c\ & 37 & 39 \\
29/07/04 & \3c\ & 35 & 34 \\
30/07/04 & \3c\ & 36 & 35 \\
31/07/04 & \3c\ & 51 & 50 \\
02/08/04 & \3c\ & 28 & 28 \\
03/08/04 & \s5\ & 26 & 27 \\
04/08/04 & \s5\ & 31 & 29 \\
05/08/04 & \s5\ & 22 & 23 \\
06/08/04 & \s5\ & 31 & 32 \\
07/08/04 & \s5\ & 28 & 29 \\
08/08/04 & \s5\ & 31 & 33 \\
TOTAL    & \s5\ & 325 & 316 \\
TOTAL    & \3c\ & 263 & 266 \\
\hline
\end{tabular}
\end{center}
\end{table}
%%%%%%%%%%%%%%%%%%%%%%%%%%%%%%%%%%%%%%%%%%%%%%%%%%%%%%%%%%%%%%%%%%%%%%%%%

%%%%%%%%%%%%%%%%%%%%%%%%%%%%%%% TABLE 2 %%%%%%%%%%%%%%%%%%%%%%%%%%%%%%
\begin{table}
\begin{center}
\caption{Reference stars in the field of \s5.}
\begin{tabular}{lcc} 
\hline
Star &    B   &    I    \\
\hline
1    &  $16.08 \pm 0.02$    &  $14.39 \pm 0.01$  \\
2    &  $16.61 \pm 0.02$    &  $14.93 \pm 0.01$  \\
3    &  $16.39 \pm 0.02$    &  $14.52 \pm 0.01$  \\
4    &  $16.78 \pm 0.03$    &  $13.24 \pm 0.01$  \\
5    &  $16.43 \pm 0.02$    &  $14.87 \pm 0.01$  \\
\hline
\end{tabular}
\end{center}
\end{table}
%%%%%%%%%%%%%%%%%%%%%%%%%%%%%%%%%%%%%%%%%%%%%%%%%%%%%%%%%%%%%%%%%%%%%%%%%

%%%%%%%%%%%%%%%%%%%%%%%%%%%%%%% TABLE 3 %%%%%%%%%%%%%%%%%%%%%%%%%%%%%%
\begin{table}
\begin{center}
\caption{Reference stars in the field of \3c.}
\begin{tabular}{lcc}
\hline
Star &    B   &    I    \\
\hline
1    &  $14.95 \pm 0.01$   &  $13.40 \pm 0.01$     \\
2    &  $15.13 \pm 0.01$   &  $13.93 \pm 0.01$     \\
3    &  $16.07 \pm 0.02$   &  $14.77 \pm 0.01$     \\
4    &  $16.76 \pm 0.03$   &  $15.13 \pm 0.02$     \\
5    &  $15.86 \pm 0.01$   &  $14.29 \pm 0.01$     \\
\hline
\end{tabular}
\end{center}
\end{table}
%%%%%%%%%%%%%%%%%%%%%%%%%%%%%%%%%%%%%%%%%%%%%%%%%%%%%%%%%%%%%%%%%%%%%%%%%

%%%%%%%%%%%%%%%%%%%%%%%%%% FIG-1 %%%%%%%%%%%%%%%%%%%%%%%%%%%%%%%%%%%%%%%%
\begin{figure}
\psfig{figure=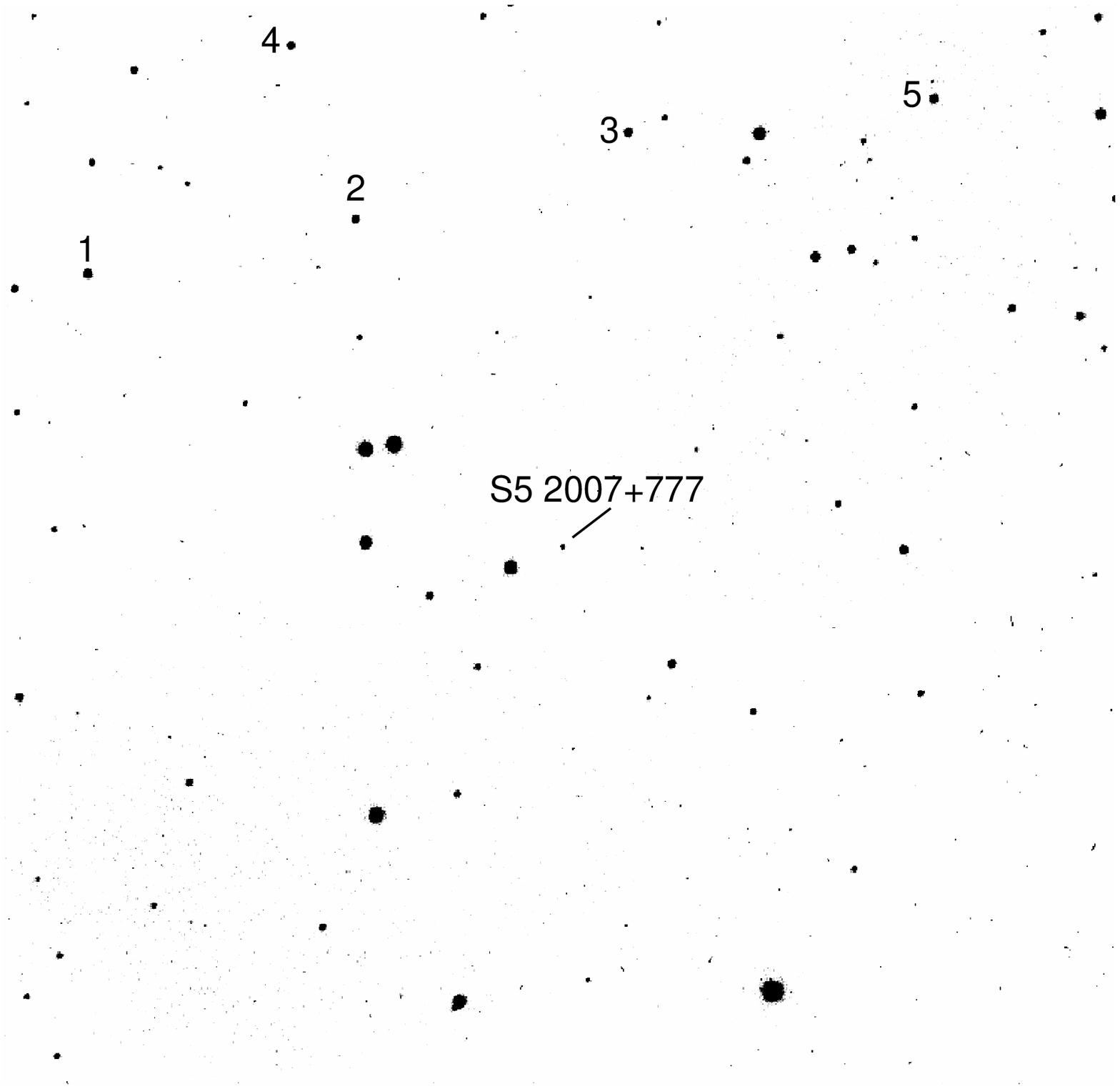,height=8.3truecm,width=8.3truecm,angle=0}
\caption[]{Finding chart ($8^{\prime}.5 \times 8^{\prime}.5$) 
of \s5 and the comparison stars. North is up
East is left.}
\end{figure}
%%%%%%%%%%%%%%%%%%%%%%%%%%%%%%%%%%%%%%%%%%%%%%%%%%%%%%%%%%%%%%%%%%
%%%%%%%%%%%%%%%%%%%%%%%%%% FIG-2 %%%%%%%%%%%%%%%%%%%%%%%%%%%%%%%%%%%%%%%%
\begin{figure}
\psfig{figure=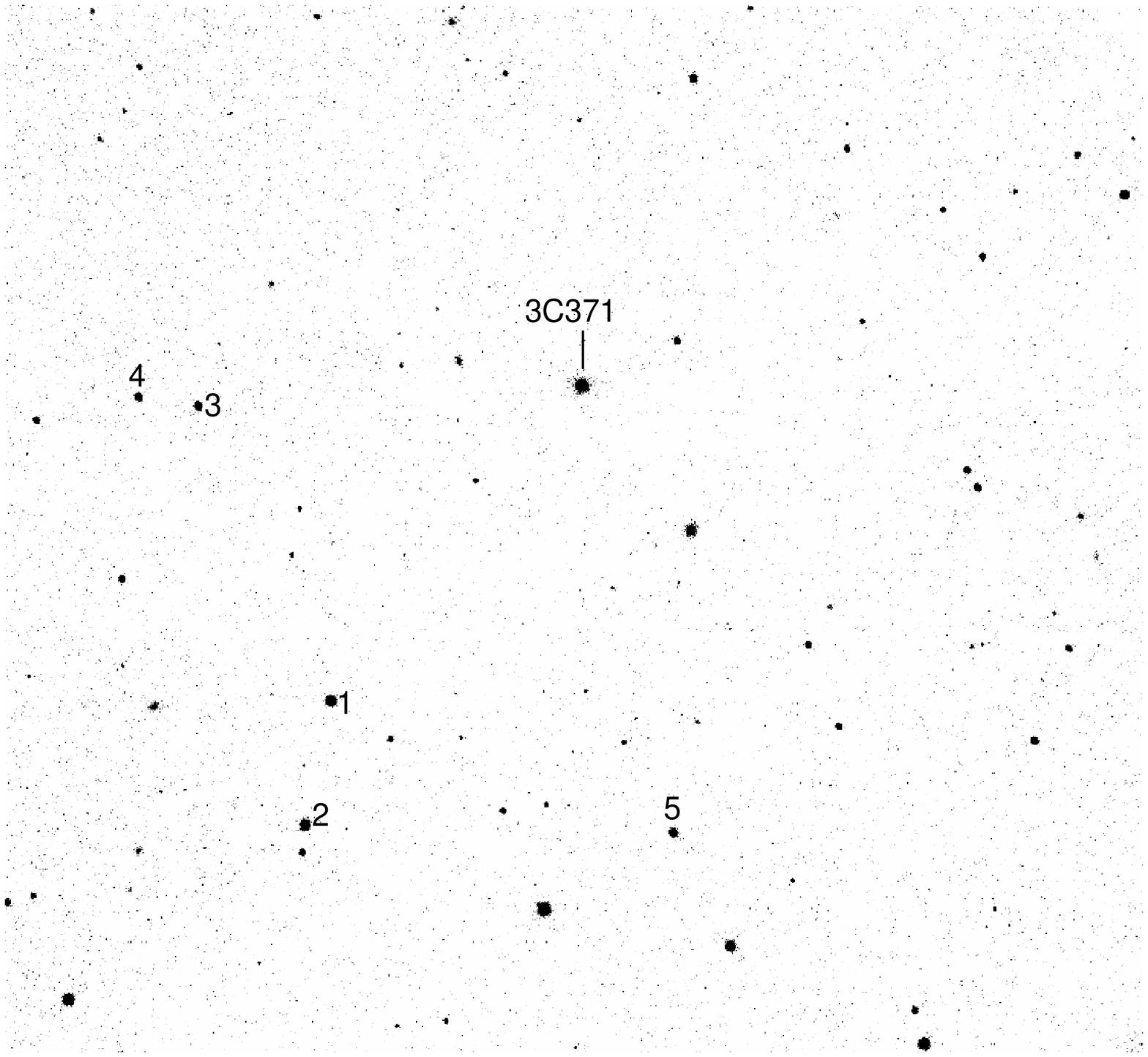,height=8.3truecm,width=8.3truecm,angle=0}
\caption[]{Finding chart ($8^{\prime}.5 \times 8^{\prime}.5$)
of \3c and the comparison stars. North is up
East is left.}
\end{figure}
%%%%%%%%%%%%%%%%%%%%%%%%%%%%%%%%%%%%%%%%%%%%%%%%%%%%%%%%%%%%%%%%%%

We did not find any published comparison star sequence in the fields of \s5 and
\3c for both $B$ and $I$ bands. We therefore chose five unsaturated stars, 
close to \s5 (see Fig. 1) and
five close to \3c (see Fig. 2) and carried out aperture photometry by
integrating counts within circular aperture centered on each star. 
Their instrumental magnitudes were transformed to the standard system through
observations of standard stars from Landolt (1992) during the nights of 5, 6
October 2001.  Tables 2 and 3 list the magnitudes of the reference stars on the
fields of \s5 and \3c respectively; quoted errors include both the uncertainty
in the determination  of the instrumental magnitudes (based on photon
statistics) and the uncertainty associated with the transformation to the
standard system. The uncertainty of the magnitude estimation of the BL Lac
objects was computed using the standard ``propagation of errors" formula (e.g.
Bevington (1969), taking into account the photometry error of the source and
standard stars' measurements together with the uncertainty in the magnitude
estimation of the comparison stars. 

The calibrated magnitudes were corrected for Galactic reddening using the vales
of $A_B = 0.696$ and $A_I = 0.313$ for \s5\ and $A_B = 0.155$ and  $A_I =
0.069$ for \3c. These vales (taken from NED\footnote{The NASA/IPAC
Extragalactic Database (NED) is operated by the Jet Propulsion Laboratory,
California Institute of Technology, under contract with the National
Aeronautics and Space Administration.}) were extracted from the Schlegel \etal\
(1998) $A_V$ maps (calculated by using the $A_{\lambda}$ versus $\lambda$
relationship of Cardelli \etal\ 1989), and are representative of the Galactic
absorption in the direction of the host galaxy. 

We converted the dereddened magnitudes into flux, and we then corrected for the
contribution of the host galaxy to the measured flux in each band as follows.
We adopted the Scarpa \etal\ (2000), and Pursimo \etal\ (2002) host galaxy,
{\it R} band magnitude estimates for \s5\ and \3c, respectively. We corrected
them for Galactic absorption, and inferred the $B_{host}$ and $I_{host}$ 
magnitudes, using the elliptical galaxy colours of $V-R = 0.90, B-V = 1.59$ and
$R-I = 0.77$ (at redshift 0.2, appropriate for \s5) and $V-R = 0.61, B-V =
0.96$ and $R-I = 0.70$ (at zero redshift, in the case of \3c) from  Fukugita
\etal\ (1995). Assuming a de Vaucouleurs $R^{1/4}$-law profile, we estimated
the host galaxy contribution within the circular aperture used for the
photometry, and subtracted it from our dereddened light curves of each object. 
Note that the subtraction of the host galaxy contribution in the case of \3c\
is necessary, otherwise this bright, ``steady-state" flux contribution can
seriously wash-out the intrinsic variability amplitude of the central source. 

%%%%%%%%%%%%%%%%%%%%%%%%%%%%%%%%%%%%%%%%%%%%%%%%%%%%
\section{The observed light curves} 
%%%%%%%%%%%%%%%%%%%%%%%%%%%%%%%%%%%%%%%%%%%%%%%%%%%%%

\subsection {S5 2007+777} 

Figs.~3, 4 and 5 show the dereddened $B$ and $I$ light curves of \s5\ during
the 2001, 2002 and 2004 observations, respectively. In the same figures we also
show the $B$ band light curve of the comparison star 3. \s5\ was in a similar
flux state during the three observing periods. The 2004 and 2002 light curves
show min-to-max variations of the order of $\sim 30-40$\% on a time scale of
$\sim 2-4$ days. During 2001, the variations are of smaller amplitude ($\sim
10$\%). 

%%%%%%%%%%%%%%%%%%%%%%%%%% FIG-3 %%%%%%%%%%%%%%%%%%%%%%%%%%%%%%%%%%%%%%%%
\begin{figure}
\psfig{figure=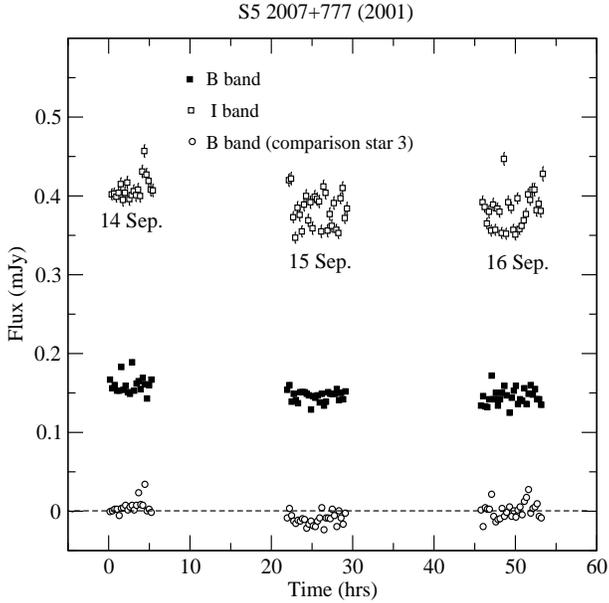,height=8.0truecm,width=8.0truecm,angle=0}
\caption[]
{$B$ and $I$ band light curves for \s5\ during the 2001 observations. Time is
measured in hours from 17:58 UT on September 14, 2001. The empty circles below
the $B$ band light curve show the $B$ band light curve of the
comparison star 3. For clarity reasons, the standard star light curve is
shifted to the mean flux level of zero mJy.}
\end{figure}
%%%%%%%%%%%%%%%%%%%%%%%%%%%%%%%%%%%%%%%%%%%%%%%%%%%%%%%%%%%%%%%%%%

%%%%%%%%%%%%%%%%%%%%%%%%%% FIG-4 %%%%%%%%%%%%%%%%%%%%%%%%%%%%%%%%%%%%%%%%
\begin{figure}
\psfig{figure=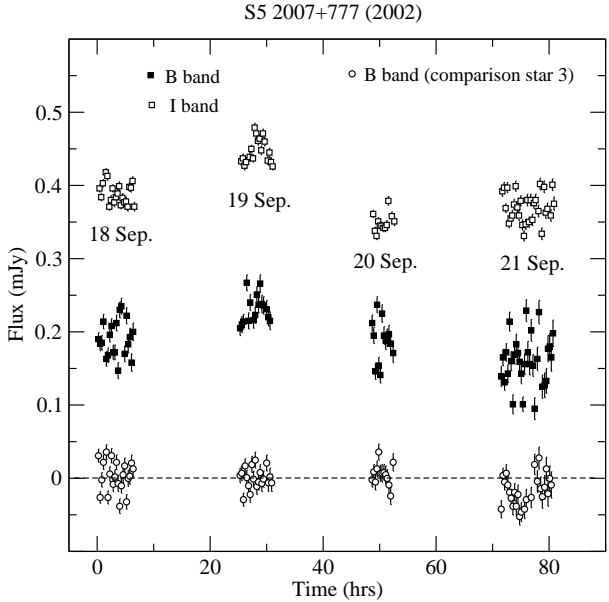,height=8.0truecm,width=8.0truecm,angle=0}
\caption[]{Same as with Fig. 3, but for the case of the 2002 observations of
\s5 (time is measured in hours from 17:50 UT on September 18, 2002).} 
\end{figure}
%%%%%%%%%%%%%%%%%%%%%%%%%%%%%%%%%%%%%%%%%%%%%%%%%%%%%%%%%%%%%%%%%%

%%%%%%%%%%%%%%%%%%%%%%%%%% FIG-5 %%%%%%%%%%%%%%%%%%%%%%%%%%%%%%%%%%%%%%%%
\begin{figure}
\psfig{figure=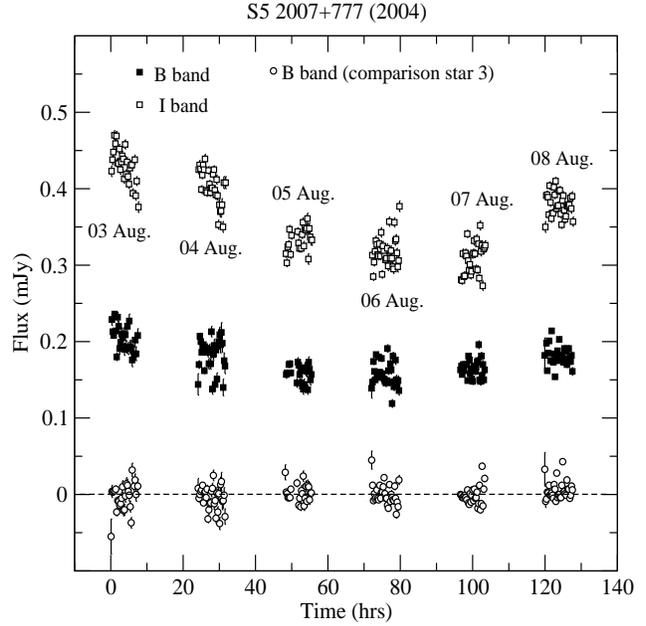,height=8.3truecm,width=8.3truecm,angle=0}
\caption[]{Same as with Fig. 3, but for the case of the 2004 observations of
\s5 (time is measured in hours from 18:12 UT on August 03, 2004).}
\end{figure}
%%%%%%%%%%%%%%%%%%%%%%%%%%%%%%%%%%%%%%%%%%%%%%%%%%%%%%%%%%%%%%%%%%

An unusual feature of the \s5\ light curves is the large intra-night 
``scatter" that is observed in both band light curves (see for example the $B$
and $I$ band 21 Sep.~2002 observations). There seems to be little correlation
between these  intra-night, low amplitude variations in the two bands. If
intrinsic, this is a rather unexpected result. However, for a source as faint
as \s5,  these variations may correspond to experimental  uncertainties in the
flux measurement of the source, which for some reason are not taken into
account by the formal errors of the photometry. In order to investigate this
issue further, we calculated the standard deviation of the $B$ and $I$ band
light curves of the comparison stars 4 and 2 during each night (these are  the
faintest of the comparison stars in the respective  bands). We find an average
scatter of $\sigma_{I}\sim 0.01-0.02$ and  $\sigma_{B}\sim 0.015-0.03$ mJy, for
the $I$ and $B$ band light curves, respectively. These are slightly  smaller
than the intra-night standard deviation of the \s5\ light curves. 

In order to investigate this issue further, we employed the techniques
described in  Howell et al. (1988). Following their prescription,
we investigated whether the very fast, small amplitude, erratic variations 
seen in the 2001 observations of \s5\ are intrinsic or not. We chose stars 2
and 4 in Table 2 as our ``C" and ``K" comparison stars. Then,  for each night,
in each band, we used the individual CCD frames to estimate $\Gamma^{2}$  on a
frame by frame basis (using their equation 13). We then calculated the average
$\Gamma^{2}_{ave}$ value, and the sample variances $s^{2}_{C-K}, s^{2}_{S5-C}$,
using the differential magnitude intra-night  light curves. The results from
the application of the test statistic for variability that these authors
suggest (equation 16 in their paper) do not confirm the $B$ band, intra-night
variations during the 2001 observations of \s5\ to be significant at a
confidence level higher than 95\%. The same holds true in other cases as well
(for example the 21 September,  2002 intra-night variations). However, we do
find that the intra-night, $B$ band variations observed during a few nights
(e.g. during the August 03 and 04, 2004 observations) are significant at the
95\% confidence level. In this case though, this result can be explained by
the  presence of well defined flux decay/rise trends rather than the scatter
around them. The picture is different in the case of the $I$ band intra night
light curves. With the exception of the September 14 2001 observations, we find
that the observed scatter in the September 15 and 16 2001 observations, the
observed small amplitude $I$ band variations are significant at the 99.9\%
confidence level. However, we note that  both the ``C" and ``K" comparison
stars are much brighter than \s5. Consequently, the average $\Gamma^{2}$ values
are significantly larger than 1, i.e. the optimum cases according to Howell
\etal\ (1988). It is not clear how this situation can affect our results, but
these authors do warn that in cases like ours one is left with a less
``strigent" test of variability.   

We  conclude that most of the observed $B$ band, intra-night, low-amplitude,
erratic variations in the \s5\ light curves are not intrinsic, while the
opposite is true in the case of the $I$ band light curves. More $I$ band
observations are required in order to verify this result, and more accurate $B$
band light curves in order to investigate whether similar variations are also
present in them. 

In order to reduce the experimental noise, we binned the light curves using
bins of size 1.5 hrs. Typically, there are $4-5$ points in each bin of that
size.  We estimated their mean and standard deviation, which we accepted as
representative of the uncertainty of the mean value in each bin. This has the
positive effect of providing an empirical estimate of the error, despite the
modest bin size, relying less heavily on theoretical calculations which may not
take into account all ``systematic" effects. 

The resulting light curves, normalized to their mean, are plotted in Figs.~ 8a,
9a and 10a and are presented in the following section. The binned light curves 
show the same variability trends, already visible in Fig.~3, 4 and 5, but more
clearly. A $\chi^{2}$ test indicates that they are all significantly variable. 

Both the $B$ and $I$ band light curves show similar variations. During August
2004,  the flux decreases slowly over a period of 4 days, and then increases in
the following two nights. Flux decreasing or increasing trends are observed
within individual nights as well (e.g. during the Aug.~3rd and 7th
observations).  During the September 2002 observations, the flux increases in
the first two nights, and then decreases in the following two nights. Finally,
during the 3 day long September 2001 observations we observe a steady, low
amplitude ($<10$\%) flux decrease in both light curves.  There is no indication
of a clear ``flare-like" event, either during individual nights or over the
whole period of each observing run. 

The fractional variability amplitude, $f_{rms}$ (defined as in Papadakis \etal,
2003) of the $B$ band 2001, 2002 and 2004 light curves is 4.3\%, 12.9\% and
9.5\%, respectively. The respective amplitude estimates for the $I$ band light
curves are: 4.1\%, 8.9\%, and 12.4\%. Within each night, the average amplitude
of the observed variations during the 2002 and 2004 observations is:
$f_{rms,B}=2.5\pm1$\%, and $f_{rms,I}=2.8\pm0.7$\%. These are comparable to the
average variability amplitude of the S4 0954+658 observations on similar time
scales, and lower than those we have measured in BL Lac, using light curves
generated by a similar sampling pattern (Papadakis \etal, 2003, 2004).  

\subsection{3C371}

Figs.~6 and 7 show the dereddened $B$ and $I$ light curves of \3c\ during the
2001 and 2004 observations, respectively. In the same figures we also show the
$B$ band light curve of comparison star 4. \3c\ was in a brighter flux state
during the 2004 observing run. 

Although \3c\ is a brighter source than \s5, considerable intra-night
``scatter" is also observed in its light curves. In this case, the standard
deviation of the comparison stars intra-night light curves is smaller than that
of the \3c\ light curves. Therefore it is rather unlikely that  the observed
scatter is caused by  unaccounted uncertainties in the photometry of the source
or of the comparison stars. However, the host galaxy of \3c\ is very bright. It
is possible then that the observed low amplitude, random variations in the
object's light curves are caused by the unavoidable variation of the  host
galaxy contribution to the measured flux in an aperture of fixed size due to
small seeing variations within each night.

%%%%%%%%%%%%%%%%%%%%%%%%%% FIG-6 %%%%%%%%%%%%%%%%%%%%%%%%%%%%%%%%%%%%%%%%
\begin{figure}
\psfig{figure=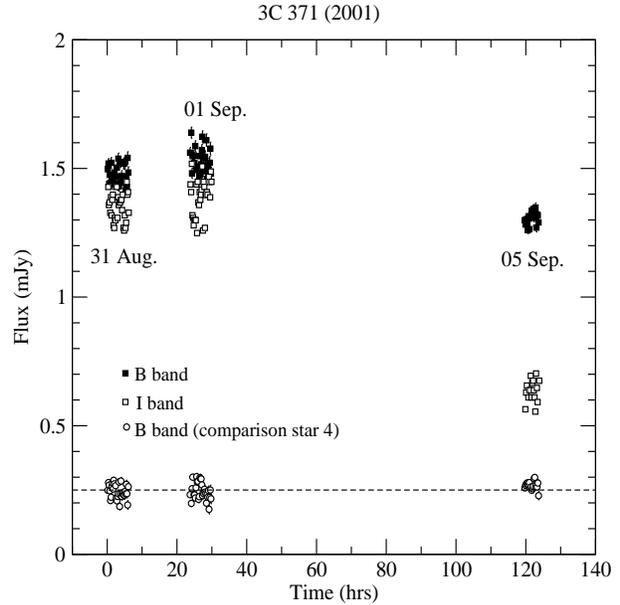,height=8.0truecm,width=8.0truecm,angle=0}
\caption[]{$B$ and $I$ band light curves of \3c\ during the 2001 observations.
Time is measured in hours from 18:16 UT on August 31, 2001. As before with
Fig.~3 in the case of \s5, the empty circles show the $B$ band light curve
of  the comparison star 4 (shifted to a mean of 0.25 mJy).}
\end{figure}
%%%%%%%%%%%%%%%%%%%%%%%%%%%%%%%%%%%%%%%%%%%%%%%%%%%%%%%%%%%%%%%%%%

%%%%%%%%%%%%%%%%%%%%%%%%%% FIG-7 %%%%%%%%%%%%%%%%%%%%%%%%%%%%%%%%%%%%%%%%
\begin{figure}
\psfig{figure=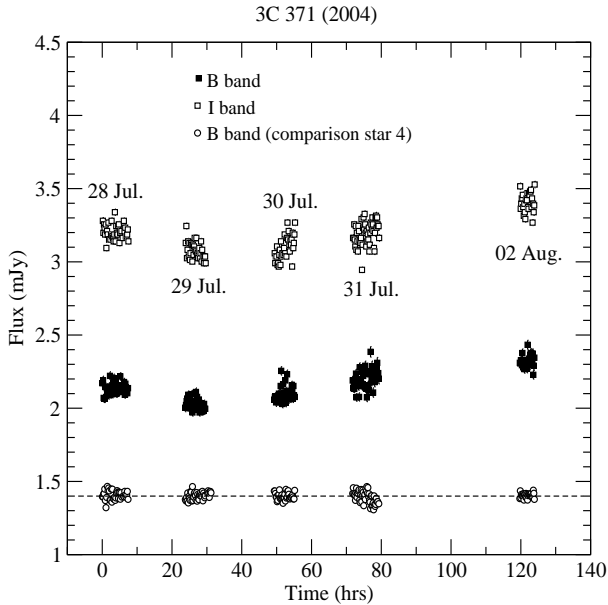,height=8.0truecm,width=8.0truecm,angle=0}
\caption[]{Same as with Fig. 6, but for the case of the 2004 observations of
\3c\ (time is measured in hours from 18:39 UT on July 28, 2004). In this
case, the comparison star $B$ band light curve is shifted to a mean of 1.4
mJy.} 
\end{figure} 
%%%%%%%%%%%%%%%%%%%%%%%%%%%%%%%%%%%%%%%%%%%%%%%%%%%%%%%%%%%%%%%%%%

As with \s5, in order to reduce the scatter in the light curves, we binned them
using bins of size 1.5 hrs. On average, there are 6 points in each bin (the 
resulting light curves, normalized to their mean, are plotted in Figs.~11a and
12a, presented in Section 4). The variability patterns evident in Figs.~6 and 7
are now more apparent in the binned light curves. 

Both band light curves show similar variations. During the July/August 2004
observations the flux decreases the first two nights, and then increases in 
the following 4 nights by $\sim 15$\%, in both bands. The behaviour of the
source is significantly different during the 2001 observations. First of all 
the $B$ band flux level is larger than that of the $I$ band (see Fig.~6). 
Then, although the flux remains almost constant during the first two nights,
the $I$ band flux drops by more than half within 4 days (no comparison star
light curve shows any sign of such a large flux drop in the same period).  This
is the largest amplitude variation  observed in the present light curves of
both objects. During the same period the $B$ band flux also decreases, but only
by a factor of $\sim 15$\%. 

The fractional variability amplitude, $f_{rms}$, of the $B$ band 2001 and 2004
light curves, taken as a whole, is 6.7\%, and 4.6\%, respectively. The
respective amplitude estimates for the $I$ band light curves are: 30.5\%, and
3.6\%. Within each night, the average amplitude of the observed variations
during the 2002 and 2004 observations is much smaller: $f_{rms,B}=1\pm0.2$\%,
and $f_{rms,I}=1.4\pm0.3$\%. This is almost half the  average variability
amplitude of \s5\ on similar time scales (i.e. of the order of a few hours).
Note that this amplitude is estimated after the subtraction of the host galaxy
contribution. Had we not subtracted it, the variability amplitude would be
significantly reduced. 

%%%%%%%%%%%%%%%%%%%%%%%%%%%%%%%%%%%%%%%%%%%%%%%%%%%%
\section{Spectral variability}
%%%%%%%%%%%%%%%%%%%%%%%%%%%%%%%%%%%%%%%%%%%%%%%%%%%%%

In order to study the spectral variability of both sources, we calculated the
ratio of the $B$ over the $I$ band flux using the 1.5 hrs binned dereddened
light curve. Assuming a power-law  like spectrum, the $B/I$ ratio is
representative of the spectrum's slope. Plots of the $B/I$ ratio light curves 
during the various observing periods are shown in   Figs.~8b, 9b, and 10b in
the case of \s5, and Figs.~11b and 12b in the case of \3c. In the lower panel
of the same figures we also plot the $B/I$ ratios versus the sum of the $B$ and
$I$ source flux, which we consider as  representative of the ``total" source
flux. These plots can be used to investigate whether the spectral variations
are correlated with corresponding flux variations or not. 

\subsection {S5 2007+777}

During the 2001 observations of \s5, the $B/I$ flux ratio stays roughly
constant at $\sim 0.38$. A $\chi^{2}$ test shows that, formally speaking, the
flux ratio is variable ($\chi^{2}_{red}=2$ for $15$ degrees of freedom, when we
fit the data with a constant). However these variations are of small amplitude
($\sim 2.5$\%), like the small amplitude flux variations we observe in the
light curves. Furthermore, as Fig.~8c shows, they  are not correlated with the
source flux, apart perhaps from the observations during September the 14th. The
data from this night are connected with a solid line in this panel. The $B/I$
ratio appears to follow a small loop-like pattern, which evolves towards the
clockwise direction as the flux evolves. 

%%%%%%%%%%%%%%%%%%%%%%%%  FIG-8 %%%%%%%%%%%%%%%%%%%%%%%%%%%%%%%%%%%%%%%%%%
\begin{figure}
\psfig{figure=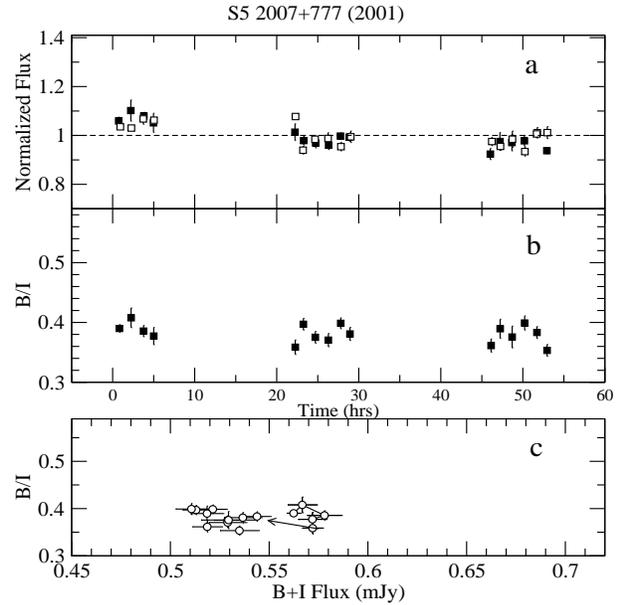,height=8.0truecm,width=8.0truecm,angle=0}
\caption[]{(a) $B$ and $I$ band light curves of \s5\ in 2001  (filled squares
and open squares, respectively), binned in 1.5 hrs  and normalized to their
average flux level. (b) The flux ratio   $B/I$ light curve. Time is measured in
hours as in Fig. 3. (c) The  $B/I$ ratio  vs.~the $B+I$ flux plot. The data of
September the 14th are connected with a  solid line and the arrow indicates
the flux evolution towards the second night of observations.}
\end{figure}
%%%%%%%%%%%%%%%%%%%%%%%%%%%%%%%%%%%%%%%%%%%%%%%%%%%%%%%%%%%%%%%%%%

%%%%%%%%%%%%%%%%%%%%%%%%  FIG-9 %%%%%%%%%%%%%%%%%%%%%%%%%%%%%%%%%%%%%%%%%%
\begin{figure}
\psfig{figure=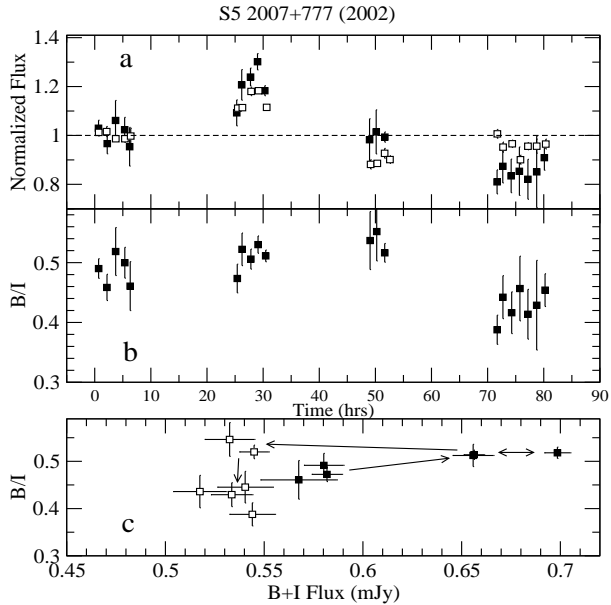,height=8.0truecm,width=8.0truecm,angle=0}
\caption[]{Same as in Fig. 8, for the 2002 observations of \s5. In the $B/I$
vs.~$B+I$ plot, we have used the 3-hrs binned light curves, for clarity reasons
(see text for details). The filled and open squares in the $B/I$ vs.~$B+I$ plot
indicate the periods when the source flux increases or decreases, respectively.
The arrows show how the source flux evolves during the observations.}
\end{figure}
%%%%%%%%%%%%%%%%%%%%%%%%%%%%%%%%%%%%%%%%%%%%%%%%%%%%%%%%%%%%%%%%%%

%%%%%%%%%%%%%%%%%%%%%%%%  FIG 10 %%%%%%%%%%%%%%%%%%%%%%%%%%%%%%%%%%%%%%%%%%
\begin{figure}
\psfig{figure=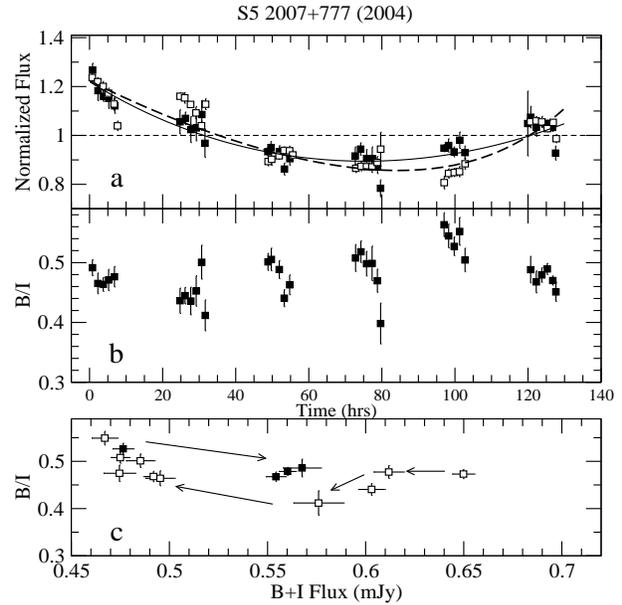,height=8.0truecm,width=8.0truecm,angle=0}
\caption[]{Same as in Fig. 8, for the 2004 observations of \s5.  The solid
and dashed lines in the upper panel show the best fitting model to the $B$ and
$I$ band light curves, respectively (see text for details). In the $B/I$
vs.~$B+I$ plot, we have used again 3-hrs binned light curves.}
\end{figure}
%%%%%%%%%%%%%%%%%%%%%%%%%%%%%%%%%%%%%%%%%%%%%%%%%%%%%%%%%%%%%%%%%%

The situation is markedly  different during the 2002 and 2004 observations. In
both cases, the larger amplitude flux variations are associated with larger
amplitude spectral variations as well: $6$\% and $6.5$\%, on average,
respectively. What is even more important though is that the spectral
variations are correlated with the source variations in these cases.  Fig.~9c
and 10c show the $B/I$ vs $B+I$ plots for the 2002 and 2004 observations. In
order to show clearly that the spectral variations are indeed correlated with
the flux changes, the $B/I$ ratios in these plots  have been calculated using
3-hrs binned $B$ and $I$ light curves. 

In the 2002 observations, as the flux rises during the first two nights 
(filled circles in Fig.~9c) the flux ratio  increases as well, implying a
``hardening" of the source's spectrum.  This ``harder when brighter" behaviour
is typical of BL Lac objects.  Then, as the flux decreases in the following
night, the flux ratio remains constant, and then decreases, although the source
flux remains roughly constant (open squares in Fig.~9c). Clearly, the spectral
slope does not evolve in the same way, i.e. it does not follow  the same path
in the $B/I$ vs $B+I$ plane during the flux rise and subsequent decay. Instead,
it follows a clear loop-like pattern, which evolves in the anti-clockwise
direction.  

We observe flux related spectral slope variations  during the 2004 observations
as well (Fig.~10c; the flux ratios shown in this plot are calculated again
using 3-hrs binned $B$ and $I$ light curves). During the first two nights, the
flux decays and  the spectrum becomes  ``softer". However, although the flux 
keeps decreasing in the following two nights, the spectrum hardens (i.e. $B/I$
increases). Finally, as the flux is rising in the last two nights of
observations, the spectrum  softens once more (filled squares in  Fig.~10c),
but  following a clearly different path than that during the original flux
decaying phase.  Although clearly related, the flux ratio does not vary with
flux in a simple linear way. Instead, the $B/I$ evolution during the flux decay
and subsequent rise defines a loop-like pattern which evolves on the clockwise
direction. 

In order to understand better the reason for the spectral variability observed
during the 2004 observations we fitted the $B$ and $I$ band light curves with
an empirical model which consists of the sum of two exponential functions, one
decaying and the other rising, i.e. $f(t)=A_{d}
e^{-t/\tau_d}+A_{r}e^{t/\tau_{r}}$, where $f(t)$ is the source flux, and
$\tau_{r,d}$ are the characteristic rising and decaying  time scales,
respectively. The best fitting curves are also shown in  Fig.~10a (solid and
dashed lines for the $B$ and $I$ band light curves, respectively). Although
they do not provide a formally acceptable fit to the observed light curves,
they describe rather well the broad flux decaying and rising trends that they
exhibit. The best fitting value of the characteristic rising and decaying time
scales are $\tau_{d,B} =86\pm10$ hrs, $\tau_{d,I}=155\pm16$ hrs, and
$\tau_{r,B}=102\pm10$ hrs,  $\tau_{r,I}=34\pm4$ hrs. 

These results provide an explanation for the spectral variations that we
observe during the 2004 observations of the source. The $B$ band flux decreases
faster at the beginning of the observations (i.e.  $\tau_{d,B}<\tau_{d,I}$). In
fact, as is evident from Fig.~10a, there seems to be a $\sim 10$ hrs delay
between the $B$ and $I$ light curve minima. Later on, although  the $I$ band
flux still decreases, the $B$ band flux remains roughly constant. 
Consequently, the $B/I$ ratio increases although the total flux still
decreases. Finally, as the the flux starts rising, the $B/I$ ratio decreases
again, because the $I$ band flux increases faster than that in the $B$ band
(i.e. $\tau_{r,B}>\tau_{r,I}$). However, since $\tau_{d,B}/\tau_{d,I} \ne
\tau_{r,B}/\tau_{r,B}$, the $B/I$ ratio does not follow the same path in the
$B/I$ vs $B+I$ plane during the flux rise and decay phases, and the loop-like
structure forms.  

\subsection {\3c}

%%%%%%%%%%%%%%%%%%%%%%%%  FIG-11 %%%%%%%%%%%%%%%%%%%%%%%%%%%%%%%%%%%%%%%%%%
\begin{figure}
\psfig{figure=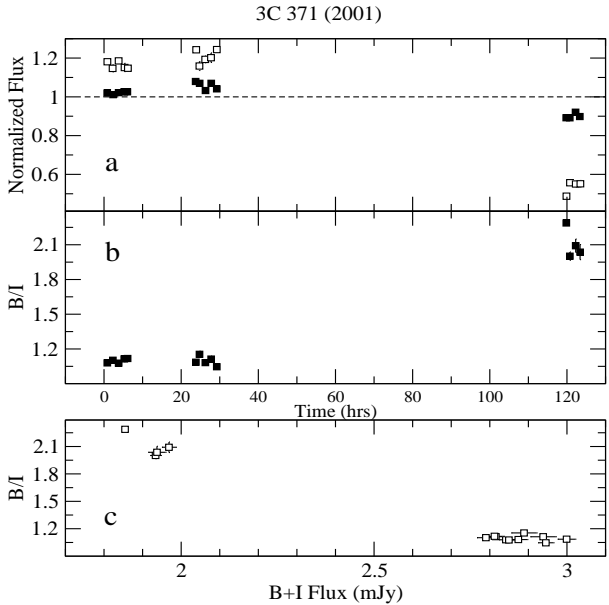,height=8.0truecm,width=8.0truecm,angle=0}
\caption[]{(a)  $B$ and $I$ band light curves of \3c\ in 2001 (filled squares
and open squares, respectively). They are binned using bins of size 1.5 hrs,
and are normalized to their average flux level. (b) In the middle panel we
plot their ratio  $B/I$ (time is measured in hours as in Fig. 9). (c)  The
$B/I$ ratio vs.~the $B+I$ flux plot.}
\end{figure}
%%%%%%%%%%%%%%%%%%%%%%%%%%%%%%%%%%%%%%%%%%%%%%%%%%%%%%%%%%%%%%%%%%

%%%%%%%%%%%%%%%%%%%%%%%%  FIG 12 %%%%%%%%%%%%%%%%%%%%%%%%%%%%%%%%%%%%%%%%%%
\begin{figure}
\psfig{figure=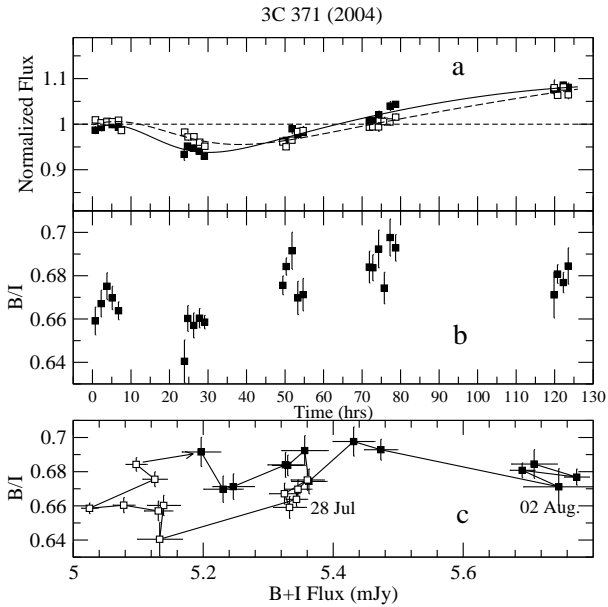,height=8.0truecm,width=8.0truecm,angle=0}
\caption[]{Same as in Fig. 6, for the 2004 observations of \3c. The
solid and dashed lines in panel (a) show the best fitting model  to the
$B$ and $I$ band light curves, respectively (see text for details). The
open and filled squares in the $B/I$ vs.~$B+I$ plot (panel c) indicate
the periods when the source flux decreases and increases, respectively.
The data of the first day of observations (28 July 2004) are marked with
the respective date. The solid line which connects the point aims at
indicating the flux and flux ratio during the following nights until
August the 2nd (last day of observations).}
\end{figure}
%%%%%%%%%%%%%%%%%%%%%%%%%%%%%%%%%%%%%%%%%%%%%%%%%%%%%%%%%%%%%%%%%%

The \3c\ 2004 observations also display flux related spectral variations. 
During the first two nights, as the flux decreases, the spectrum becomes softer
(i.e. $B/I$  decreases) as well. In the following three nights, as the flux
rises  the $B/I$ ratio increases as well. The flux ratio does not follow a
well defined linear trend in either case. Instead, a loop-like pattern which
evolves in the clock-wise direction is visible in the $B/I$ vs $B+I$ plot of
these observations (Fig.~12c). 

In Fig.~12a, together with the normalised, binned $B$ and $I$ light curves, we
also show two lines which describe rather well the flux evolution in the two
bands during our observations (solid and dashed lines, for the $B$ and $I$
light curves, respectively). These lines correspond to the best fitting curves
of a model which consists of the sum of two Gaussians, i.e.
$f(t)=(A_{1}/\sigma_{1})exp\{-(t-t_{m,1})^2/\sigma^2_{1}\}+
(A_{2}/\sigma_{2})exp\{-(t-t_{m,2})^2/\sigma^2_{2}\}$, where $f(t)$ is the
source flux  (exponentially decaying and rising functions  do not describe well
the observed flux variations in this case). The rate of the flux variations is
characterised by the width of the Gaussians; the wider the Gaussian (i.e. the
larger the $\sigma$) the slower the flux varies. Our best fitting $\sigma$
values are: $\sigma_{1,B}=14.1\pm1.5, \sigma_{2,B}=196\pm16,
\sigma_{1,I}=18\pm2,$ and $\sigma_{2,I}=317\pm25$ hrs. These results suggest
that the $B$ band decays and rises faster than the $I$ band light curve. Again,
as in the case of the \s5\ observations in 2004, there appears to be a $\sim
10-12$ hrs delay between the $B$ and $I$ light curves at minimum (with the $B$
band light curve leading).  

As a result of the differences between the $B$ and $I$ rise and decay time
scales, the $B/I$ ratio decreases as the flux decays, however, as it
subsequently rises, the flux ratio increases because $B$ is rising faster than
$I$. Only during the last night of observations, when the $B$ light curve may
have started to decay while $I$ is still rising, the flux ratio appears to
decrease again. 

Finally, during the first two nights of the 2001 observing run of \3c, the
$B/I$ ratio remains constant at $\sim 1.1-1.2$, a value already larger than the
flux ratio during the 2004 observations. Then, in the last night the flux ratio
increases to $\sim 2.1$, implying a remarkable spectral hardening despite the
fact that the source flux decreases. As we mentioned in the previous section,
the optical flux variability behaviour of the source during these observations
is rather unusual for a BL Lac object. 

%%%%%%%%%%%%%%%%%%%%%%%%%%%%%%%%%%%%%%%%%%%%%%%%%%%%%%%%
\section{Cross correlation analysis}
%%%%%%%%%%%%%%%%%%%%%%%%%%%%%%%%%%%%%%%%%%%%%%%%%%%%%%%%%%

In general, the presence of flux related spectral variations imply the
existence of delays between the variations in different bands, caused by the
different characteristic time scales on which they evolve (as we showed in the
previous Section). In order to investigate this issue further, we estimated the
cross-correlation function (CCF) using the ``Discrete Correlation Function" of
Edelson \& Krolik (1988). We used the original light curves (i.e. without any
binning),  and estimated the CCF for the observations of \s5\ in 2002 and 2004
observations, and for the observations of \3c\ in 2004, up to lags of $\pm4$
hrs (a time scale which is roughly equal to half the duration of our
observations each night). In all cases we used a  lag of size 0.5 hrs. The CCFs
are estimated in such a way so that a positive lag would imply that the $B$
band leads the $I$ band variations. 

Our results are shown in Fig.~13. The two \s5\ CCFs look very similar. For that
reason we added them together to estimate a mean \s5\ CCF, and this is what we 
plot in Fig.~13.  Both the \s5\ and the \3c\ CCFs show maxima of the order of
$\sim 0.75-8$. Furthermore, they are very broad, with no clear indication of
any significant delays between the the $B$ and $I$ band variations. The large
CCF maxima are in agreement  with the fact that the $B$ and $I$ light curves of
the two objects  show almost identical variations within each individual night.
The lack of any delays up indicates that the $B$ and $I$ band variations, on
time scales of minutes/hours, happen almost simultaneously. 

In the previous Section, based on the model fitting of the observed light
curves, we found evidence of delays between the $B$ and $I$ band variations of
the order of $\sim 10$ hrs, a time scale much longer than the lags on which we
have estimated the CCFs shown in Fig.~13. Although the length of some of our
light curves is considerably longer than 10 hrs, the CCF estimation on these
long time scales is not easy, due to the presence of long gaps in them.
Nevertheless, in the lower panel of Fig.~13 we plot the CCF of the 2004 \s5\
observations, estimated up to lags of $\pm 50$ hrs, using a lag bin of size 4.5
hrs. This is the best of the available light curves for the estimation of the
CCF on long time scales, because it has the largest number of consecutive 
nights with available observations.

%%%%%%%%%%%%%%%%%%%%%%%%  FIG 13 %%%%%%%%%%%%%%%%%%%%%%%%%%%%%%%%%%%%%%%%%%
\begin{figure}
\psfig{figure=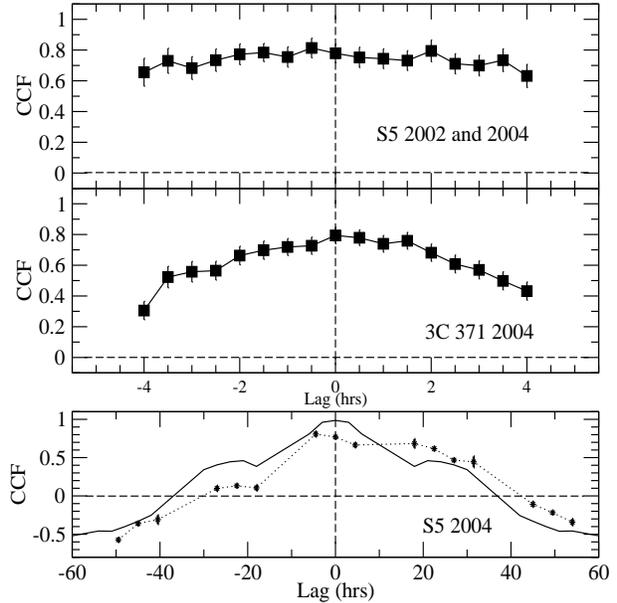,height=8.0truecm,width=8.0truecm,angle=0}
\caption[]{CCF between the $B$ and $I$ band light curves of the 2002 and 2004 
\s5\ and of the 2004 \3c\ light curves (upper and middle panels, respectively).
The points plotted with stars in the lower panel show the 2004 \s5\ CCF,
estimated using a lag size of 4.5 hrs, up to lags $\pm 50$ hrs. The solid line
in the same plot shows the auto-correlation function of the $I$ band light
curve.} 
\end{figure}
%%%%%%%%%%%%%%%%%%%%%%%%%%%%%%%%%%%%%%%%%%%%%%%%%%%%%%%%%%%%%%%%%%

The CCF is not well defined, but there appears a clear asymmetry towards
positive lags. This can be easily seen if one compares the observed CCF with
the $I$ band auto-correlation function (ACF), which is also plotted in the
same panel (solid line). At negative lags, and at lags $\le 10$ hrs, the CCF
points are consistently below the ACF, while the opposite is true at larger
lags. This result implies that, at lags longer than $\sim 10$ hrs, the
correlation of the $I$ band with the $B$ band light curve is stronger than
that with itself (and vice versa for smaller and negative lags). In order to
get an estimate of the average delay between the two light curves we fitted
the observed CCF between lags $-30$ and $30$ hrs with a Gaussian. We find a
delay of $8.6\pm2$ hrs, (with the $B$ band leading), comparable to the delay
in the $B$ and $I$ light curves minima that we had found in the previous
section.

We conclude that, despite the low intra-night variability amplitude, the $B$
and $I$ light curves of \s5\ and \3c\ are highly correlated on time scales of
the order of a few hours. Any delays must be smaller than $\sim 1$ hour. On
longer time scales though, at least during the \s5\ 2004 observations, there is
an indications that the variations in the  $B$ band light curve are leading
those in the $I$ band light curve  by $\sim 8$ hrs.

%%%%%%%%%%%%%%%%%%%%%%%%%%%%%%%%%%%%%%%%%%%%%%%%%%%%%%%%%%%%%%%%%%%%%%%
\section{Discussion and conclusions}
%%%%%%%%%%%%%%%%%%%%%%%%%%%%%%%%%%%%%%%%%%%%%%%%%%%%%%%%%%%%%%%%%%%%%%%%

We have observed \s5\ and \3c\ in the $B$ and $I$ bands for 14 and 8 nights,
respectively, during various periods in 2001, 2002 and 2004. The light curves
last for $\sim 6-9$ hours. There are $15-30$ points in each of them, almost
evenly spaced, with an average sampling interval of $\sim 0.15-0.3$ hours, on
average. These are the most detailed short term, optical monitoring
observations in two bands of these two objects to date. Our results can be
summarized as follows:

1) Both sources shows intrinsic, low amplitude variations ($\sim 2.5$\%  and
$1-1.5$\% in the case of \s5, and \3c\ respectively) on time scales as short as
a few hours. On time scales of a few days we observe broad flux rising or
decaying trends with an average amplitude of $5-12$\% (\s5) and $4-6$\% (\3c).
In the case of the 2001, $I$ band light curve of \3c\  we observed a large
amplitude variation, of the order of $70$\%, within 5 days. 

2) The observed small amplitude intra-night  variations in the two bands are
highly correlated, with no delay larger than $\sim 1$ hr detected in any
case. 

3) The 2004 \3c\ and \s5\ observations resulted in the longest and best sampled
light curves. They show well defined broad decaying and rising flux patterns.
We find that the flux rising and decaying time scales are not equal, neither
within or between the two band light curves. In both objects, we also observe a
$\sim 10-12$ hrs delay between the flux minima in the $B$ and $I$ light curves.
In the case of \s5, this result is also supported by the CCF analysis, which
reveals a similar delay between the $B$ and $I$ band observations. 

4) We observe spectral variations in almost all cases. They are clearly
correlated with the flux variations in the case of the 2002 and 2004 \s5, and
2004 \3c\ observations. We have observed the usual ``harder when
brighter/softer when fainter" behaviour (during the \s5\  observations in 2002
and 2004, respectively) but we have also observed ``unusual" behaviour like
spectral softening while the flux increases (during the \s5\ observations in
2004) and the significant hardening of the spectrum during a large flux decrease
in the 2001 \3c\ observations. 

5) In general, the relation between flux and spectral variations is not linear.
Instead, we have detected loop-like patterns in the ``spectral slope vs flux"
plane which evolve in the clockwise (during \s5\ and \3c\ 2004 observations)
and anti-clockwise direction (during the \s5\ 2002 observations). 

Variability observations of BL lac objects in more than one bands can in
principle impose significant constraints on the mechanism that causes the
observed variations in the source. To this end, we briefly investigate below
some consequences of our results from the 2004 observations of both objects
mainly.  

The continuum emission in BL Lacs is believed to arise from a relativistic jet
oriented close to the observer, i.e. the emitted radiation is highly beamed in
the forward direction. One possible mechanism that can explain the observed
variations in some cases is if the optical emission is produced in discrete
blobs moving along magnetic fields, and the viewing angle (i.e. the angle
between the blob velocity vector and the line of sight) varies with time. In
this case the beaming or Doppler factor should vary accordingly and, since the
observed flux depends on this, viewing angle variations can result in flux
variations (see e.g. Wagner  \etal\ (1993) in the case of S4 0954+658 optical
variations). A change of less than 0.5 degrees only can explain flux variations
of the order of $20$\% (Ghisellini \etal\ 1997). In this case,  we would expect
the variations to be ``achromatic" (assuming no spectral break within the
considered optical bands). However, during the present observations, we do
observe significant spectral variations which are correlated with the
associated flux variations. Therefore, we conclude that the observed
variability  in the light curves of these sources do not have a geometric
origin.

On the other hand,  the significant, flux related, energy spectral variations
that we observe can be explained if a perturbation activates a region in the
jet, filled with energetic particles  which emit through synchrotron radiation.
In this case, the important time scales which govern the spectral evolution of
the source are the acceleration, cooling, escape and light travel time scales. 

The first obvious constrain that our results can impose on this class of models
stems from the fact that the decay and rising time scales are not equal during
the 2004 observations of both objects. The conclusion is that the flux evolution
is not determined by the light source crossing time. If both the particle
injection and cooling process were operating on time-scales shorter than $R/c$
(where $R$ is the source radius) we should expect the flux rise and decay to be
symmetric. Therefore, either the injection/acceleration or/and the cooling time
scales must affect the observed variations. 

We find that during the first part of the 2004 \s5\ and \3c\ light curves, the
flux decays faster in the $B$ band, hence the time difference between the
minimum in the  $B$ and $I$  band light curves. This could be explained by the
cooling of the emitting particles due to synchrotron and self-Compton radiative
losses. For a particle of energy $\gamma mc^2$, emitting at an observed
frequency of $\nu_o\propto \gamma^2$, the cooling time is $\propto
\nu_o^{-0.5}$. The ratio of the $B$ and $I$ band median frequencies is roughly
1.8, implying that $t_{cool,B}/t_{cool,I}\propto (\nu_B/\nu_I)^{-0.5}\sim 0.7$.
This value is very close to the ratio  of $\tau_{d,B}/\tau_{d,I}=0.6\pm0.1$
that we estimated in section 3.1, in the case of the 2004 \s5\ data, and to the
ratio of $\sigma_{1,B}/sigma_{1,I}=0.8\pm0.1$ in the case of the \3c\ light
curves. It is indeed possible then, that the flux evolution during the first
part of these observations is governed by the radiative losses of the emitting
particles. 

The spectral evolution is different in the second part of the \s5\ and \3c\
light curves, when the flux starts rising. In \s5,  we observe the $I$ band
flux rising faster than the flux in the $B$ band. This situation could
correspond to the initial stages of the passage of a shock front in the
emitting region. In particular, the decrease of the flux ratio with increasing
flux that we observe during the flux rising phase  (filled squares
of Fig.~10c) is expected in the case when the cooling and acceleration
time scales of the emitting particles are comparable. In this case, particles
are gradually accelerated into the radiating window, and the information about
the onset of a flare initially propagates from lower to higher energy (Kirk, 
Rieger \& Mastichiadis 1998). Indeed, the spectral slope vs flux evolution that
we observe  is almost identical to the spectral evolution during the initial
phase of the flare as shown in Fig.~4 of Kirk et al.(1998).

On the other hand, during the flux rising phase of the 2004 \3c\ light curves,
we observe the opposite effect: the flux rising first in the $B$ and then in 
the $I$ band. This behaviour is expected in the case when the source is
injected with particles at high energies for a long period (which could be
comparable to $R/c$). In this case, the emission starts increasing at high
energies (i.e. in the $B$ band) for as long as the injection continues and the
emitting volume increases accordingly. The particles start emitting at lower
frequencies (i.e. in the $I$ band) only after some time equal to the cooling
time scale,  hence the delay in the increase of the flux in the $I$ band. 

The unusual behaviour of \3c\ during the 2001 observations is difficult to
explain. The fact that $B_{flux}>I_{flux}$ can result in the case when the peak
synchrotron emission frequency has moved above the optical band and the optical
spectral slope is larger than $-1$. However, this peak emission frequency
increase is usually caused by the onset of a major jet perturbation which also
produces a significant flux increase. Since the source flux during the 2001
observations is lower than that in the 2004 observations, this is probably not
a valid scenario. The opposite effect, i.e. a significant decrease in the peak
emission frequency, is another possibility. Perhaps the $B$ band is already 
dominated by the Compton component (produced by particles of small energy,
whose cooling time is long) at the beginning of the observations while the 
sudden, large amplitude flux drop in the $I$ band between 1st and 5th of
September, marks the transition of the $I$ band flux from the synchrotron to
the Compton regime. However, due to the sparse sampling of the available light
curves no further conclusions can be drawn. 

The observed spectral and flux variations of \3c\ and \s5\ are similar to those
observed by us in BL Lac and S4 0954+658, using multicolour light curves of
similar length and sampling pattern. All these four objects are ``Low-Frequency
peaked BL lacs" (LFBs), i.e. the synchrotron peak in their spectrum is located
at mm/IR/optical wavelengths (see e.g. Peng \etal\ 2000 in the case of \s5, and
Sambruna. Maraschi \& Urry, 1996, in the case of \3c). Their optical emission
corresponds to a frequency range around and above the synchrotron flux maximum,
and should be emitted by the most energetic particles with the in the jet.
Their cooling time scales can be quite small, and, depending on the
acceleration time scale and the size of the source, multi-band  optical
monitoring of these sources  could be particularly beneficial as it can reveal
all sorts of complicated flux related spectral variations, even on time scales
of minutes/hours. 

Our results offer direct evidence that the fast, low amplitude variations in 
the optical light curve of these 4 LBLs are caused by rapid perturbations in
the jet, which evolve on time scales of hours/days (except from S4 0954+658
where some of the short term variations could have a geometric origin). As
shock fronts overrun different regions in the jet, we expect to observe flux
related energy spectral variations, as we do.  The fact that we do not observe
the same variability behavior at all times within each source, or between
different sources, implies that  these perturbations do not always evolve on
the same time scale, and they affect regions of different size in the jet. 

\begin{acknowledgements}
We thank the anonymous referee for useful comments.
Skinakas Observatory is a collaborative project of the University of Crete, the
Foundation for Research and Technology-Hellas, and the Max--Planck--Institut
f\"ur extraterrestrische Physik. Part of this work was supported by the
European Commission under the TMR program, contract number HPRN-CT-2002-00321.
\end{acknowledgements}

\end{document}